\documentclass[journal=cmatex,manuscript=article]{achemso}

\usepackage[version=3]{mhchem}
\usepackage{amssymb}
\usepackage{amsmath}
\usepackage{graphicx}
\usepackage{multirow}

\title{Cobalt Substitution in CuFe$_{2}$O$_{4}$ spinel and its influence on the crystal structure and phonons}

\author{M. D. P. Silva}
\author{F. C. Silva}
\affiliation[Departamento de Qu\'{\i}mica, CCET, Universidade Federal do Maranh\~ao, 65085-580, S\~ao Lu\'{\i}s-MA, Brazil]{Universidade Federal do Maranh\~ao}
\author{F. S. M. Sinfrônio}
\affiliation[Departamento de Engenharia El\'etrica, CCET, Universidade Federal do Maranh\~ao, 65085-580, S\~ao Lu\'{\i}s-MA, Brazil]{Universidade Federal do Maranh\~ao}
\author{Alexandre Rocha Paschoal}
\affiliation[Departamento de F\'{i}sica, Universidade Federal do Cear\'a, Campus do Pici, 60455-760, Fortaleza - CE, Brazil]{Universidade Federal do Cear\'a}
\author{E. N. Silva}
\affiliation[Departamento de F\'{\i}sica, CCET, Universidade Federal do Maranh\~ao, 65085-580, S\~ao Lu\'{\i}s - MA, Brazil]{Universidade Federal do Maranh\~ao}
\author{Carlos William de Araujo Paschoal}
\altaffiliation{Author whose correspondence should be addressed: Phone +55 98 3301.9209; Fax: +55 98 3301 8204}
\affiliation{Departamento de F\'{\i}sica, CCET, Universidade Federal do Maranh\~ao, 65085-580, S\~ao Lu\'{\i}s - MA, Brazil}
\affiliation{Department of Materials Science and Engineering, University of California Berkeley, 94720-1760, Berkeley - CA, United States}
\email{paschoal@ufma.br}
\email{paschoal.william@berkeley.edu}
\email{paschoal.william@gmail.com}

\begin{document}

\begin{abstract}
In this work nanometric spinel Co${}_{1-x}$Cu${}_{x}$Fe${}_{2}$O${}_{4}$ powders were obtained by polymeric precursors method at several annealing temperatures between 700 and 1200 ${}^\circ$C. The samples were characterized by means of X-ray powder diffraction, confirming the ideal inverse spinel structure for CoFe${}_{2}$O${}_{4}$ sample and the tetragonal distorted inverse spinel structure for CuFe${}_{2}$O${}_{4}$ sample. Based on FWHM evaluation, we estimated that crystallite sizes varies between 27 and 37 nm for the non-substituted samples. The optical-active modes were determined by infrared and Raman spectroscopies. The phonon spectra showed a local tetragonal distortion for mixed samples.

\end{abstract}

\section{Introduction}

Multiferroics  are  singular  materials  that  exhibit  simultaneously two ferroic orders, usually electric and magnetic,  with  a  magnetoelectric coupling  between  them in  some  cases.  These materials have attracted much attention due to their potential technological applications. As such, multiferroic compounds  bring  new  functionalities  to  spintronics and  new  device  possibilities,  such  as memories,  sensors,  actuators  and  tunable  filters \cite{CHEONG2007,CHIBA2008,EERENSTEIN2007,RAMESH2007,SCOTT2007}. The most investigated multiferroic compound is the perovskite BiFeO${}_{3}$ \cite{BEA2009A,CHU2008A,SEIDEL2009,YANG2009,ZHAO2006,AGUILAR2009,BEA2008,KORNEV2007,WANG2003}. However, the long search for the control of electrical properties by magnetic fields has been recently  led to in a group of materials known as "frustrated magnets", such as: the perovskites REMnO${}_{3 }$\cite{KIMURA2003,LEE2008,PIMENOV2009} and REMn${}_{2}$O${}_{5}$ \cite{BODENTHIN2008,HARRIS2008,OKAMOTO2007,SUSHKOV2007} (with RE being rare earth ions), in the spinels CuFeO${}_{2}$ \cite{WANG2008,YE2007} and CoCr${}_{2}$O${}_{4}$ \cite{CHOI2009}, and the Y-type hexagonal ferrites (Ba, Sr)${}_{2}$Zn${}_{2}$Fe${}_{12}$O${}_{22}$ \cite{KIMURA2005}, among others.

The restricted number of pure multiferroic compounds is a consequence of the atomic arrangement specificity, since, theoretically, only 13 point groups can lead to the multiferroic behavior. In addition, by definition, ferroelectrics are insulators (3d transition metal oxides typically have ions which have an electronic distribution d$^0$), while ferromagnets need conduction electrons. The difficulties associated with the combination of electric and magnetic responses in a single phase compound can be solved by making two-phase multiferroic composites consisting of a ferroelectric component (e.g. PbZr$_{1-x}$Ti$_x$O$_3$  - PZT) and a ferromagnetic component (e.g. CoFe$_2$O$_4$) \cite{Li2007,Wang2012,Yan2011,Wang2010,Chen2010,Zhang2010,Zhang2009,Yan2009,Wu2009,Pang2009,Chen2009,Dix2007,Thang2008}. In such composites, the magnetoelectric effect is due to the interaction between elastic components of ferroelectric and ferromagnetic constituents. In such case, an electric field induces a voltage in the ferroelectric which is transferred to the ferromagnet, that causes the magnetization of the material. Since the magnetoelectric effect is higher if the coupling at the interface is higher, compounds with large surface areas (as in multilayer thin films and nanometric powders) and strongly ferroelastic are particularly interesting. This approach opens new routes to get a magnetoelectric response with the specific choice, relationship and microstructure of the constituents. In fact, the magnetoelectric coefficients at room temperature which was achieved exceed those found in low-temperature single phase samples in three to five orders of magnitude. Thus, magnetoelectric multiferroic composites are on the boundary of the technological applications \cite{EERENSTEIN2007,RAMESH2007,EDERER2005,TOKUNAGA2009}. So, as the spinels have been continuously employed to obtain good multiferroic composites, a careful attention has been given in last years to the synthesis of spinels in last years for different methods \cite{Sun2012,Tang2012,Pui2011,Girgis2011,Fei2011,Naseri2011,El-Okr2011,Gatelyte2011,Yu2011,Gatelyte2011b,Shen2010,Zhang2010b,Ajroudi2010,Veverka2010,Sertkol2010,Shahane2010,Hou2010,Shen2010b,MANOVA2009,Gozuak2009,Maaz2009}.

Besides their extensively investigated ferromagnetic properties, Li-based spinels has been extensively investigated to be applied as electrode in batteries \cite{Cabana2012,Lee2012,Lee2012a,Qian2012,Shin2012}. Also, binary spinel ferrites like CdFe$_2$O$_4$, NiFe$_2$O$_4$, ZnFe$_2$O$_4$ and CuFe$_2$O$_4$ also show significant gas-sensing activity \cite{SUN2007}. Moreover, CuFe$_2$O$_4$ shows high electronic conductivity, high thermal stability and high activity as green anode for aluminum electrolysis \cite{SELVAN2003}. Another important binary spinel-type ferrite is the cobalt ferrite CoFe$_2$O$_4$ due to its high cubic magnetocrystalline anisotropy, high coercivity, and moderate saturation magnetization, being such properties severely affected by to the concentration of divalent metal cation \cite{MANOVA2009,Lima2012,Lopez-Dominguez2013,Pereira2012,Tuysuz2012}.

Usually CoFe$_2$O$_4$ crystallizes in a spinel structure. The cationic distribution in the spinel structure can be described by the chemical formula $($Fe$_{1-x}$Co$_x)$ $[$Fe$_{1+x}$Co$_{1-x}]$O$_4$, where (  )  and  [  ] denote the tetrahedral $A$ and octahedral $B$ sites, respectively. The inversion parameter $x$ is equal to 0 for inverse spinels and to 1, when the spinel is normal. CoFe$_2$O$_4$ spinel structure is  predominantly inverse $(x = 0)$, with Co$^{2+}$ ions occupying mainly B sites while Fe$^{3+}$ ions are distributed almost equally on $A$ and $B$ sites. The inversion index $(1 -x)$ depends on the thermal history of the sample \cite{SAWATSKY1969}. By the other side, CuFe$_2$O$_4$ has a tetragonally deformed spinel structure, that is stretched along the  $<011>$ direction \cite{NEDKOV2006}. Yokoyama {\it et al.} \cite{YOKOYAMA1998} observed changes in the crystal structure of nanosized CuFe$_2$O$_4$ powders obtained by coprecipitation followed by annealing. They observed that the copper spinel is cubic at temperatures below 300 $^o$C and tetrahedral over 400 $^o$C. The formation of considerable quantities of Cu$^+$ in the lattice is the mechanism responsible for the transition from tetragonal to cubic structure \cite{YOKOYAMA1998}. As showed by Kester et al. \cite{KESTER1997}, through reduction reaction of quenched samples of CuFe$_2$O$_4$, the formation of Cu$^+$takes place, but its fraction in the  B-sublattice strongly depends on the synthesis procedures and on the subsequent temperature treatment. Nanosized particles of CuFe$_2$O$_4$, obtained by a classical ceramic technology, have also been studied \cite{JIANG1999}. High temperature treatments lead to structural and magnetic surface disorders, which can be induced by the dispersion of different copper ions in the sublattices, by the arising of cations and oxygen vacancies, by the structure amorphisation, among others. Also, when the spinel is synthesized using classical solid state route with accurate stoichiometry $(x = 1)$, it has a tetragonal structure \cite{HAMDEH1997}.

However, independently of the cation present in the lattice, the physicochemical properties of such spinel-type ferrites depend on their microstructural properties, which are related, in turn, to the preparation methods of these compounds. Thus, several routes were used to produce these binary oxides such as hydrolysis \cite{RAO2001}, ball-milling, solid state reaction, co-precipitation and sol gel methods, combustion processing etc. \cite{TAO2000}. Since the synthesis route is crucial over the spectroscopic and structural properties of the spinel-type ferrites, the aim of this work was to evaluate the effect of cobalt isovalent substitution in  CuFe$_2$O$_4$ spinels nanoparticles obtained by polymeric precursors method.

\section{Materials and Methods }

All samples were synthesized using the Polymeric Precursor Method, as follow: ferric chloride hexahydrate (FeCl${}_{3}$.6H${}_{2}$O, Isofar), copper sulphate pentahydrate (CuSO${}_{4}$.5H${}_{2}$O, Isofar) and cobalt chloride hexahydrate (CoCl${}_{2}$.6H${}_{2}$O, Vetec) were used as purchased without further purification (pa purity). The precursor solutions of Fe, Cu and Co were prepared by adding the raw solids into an aqueous solution of citric acid (C$_6$H$_8$O$_7$, Nuclear), using stoichiometric quantities. Therefore, ethylene glycol (C$_2$H$_6$O$_2$, Nuclear) was added to the metallic solution, according to a molar ratio 1:3. This solution was heated at 110 $^\circ$C for 5 h in an oven to promote polymerization. Soon after, the polymerized gel was heated at 300 ${}^\circ$C for 1 h, under air atmosphere, to burn the organic matter and form a black solid mass (primary calcination). Such carbonaceous mass was grounded until its particles were 100 Mesh sized and heat-treated at 300 ${}^\circ$C for 12 hours (secondary calcination), under a high oxygen atmosphere, in order to produce oxygen vacancies in the solids. Finally, such precursor powders were annealed between 700--1200 ${}^\circ$C (ternary calcination) in air for 4 h  in Al${}_{2}$O${}_{3}$ crucibles, and the desired spinel compounds resulted.

Crystal structures of the annealed powders were examined using an X-ray diffractometer Xpert MPD (Panalytical), with Co K$\alpha$ radiation (40 kV and 40 mA), speed of 0.02$^{\circ}$ $\theta s^{-1}$ and value ranging from 10 to 100 $^\circ$. X-ray powder diffraction (XRD) patterns were compared with the Joint Committee Powder Diffraction Standards (JCPDS) data for the phase evaluation.

Crystallite sizes $\left( D \right)$ of the samples were determined from X-ray line broadening analyzes, employing the Scherrer's equation:
\begin{equation}\label{scherrer}
  D = \frac{K\lambda}{\beta cos\theta}
\end{equation}
where $\lambda$ is the X-ray radiation wavelength ($\lambda$ =1.78896 \AA), $K$ is the Scherrer constant, $\beta$ is the FWHM of the peak (in radians) and $\theta$ is the peak angular position.

The infrared spectra were obtained using an IR prestige-21 infrared spectrometer (Shimadzu), applying KBr as dispersant agent (1:100 wt./wt.) in the mid range: 400 up to 1000 cm$^{-1}$.

The confocal Raman spectra were acquired using an alpha 300 system microscope (Witec, Ulm, Germany), equipped with a highly linear (0.02\%) stage, piezo-driven, and an objective lens from Nikon (20x, NA = 0.40). A Nd:YAG polarized laser ($\lambda $ = 532 nm) was focused with a diffraction-limited spot size (0.61$\lambda $/NA) and the Raman light was detected by a high sensitivity, back illuminated spectroscopic CCD behind a 600 g/mm grating. The final power at the end of the objective lens used to focus on the sample was 3 mW. The spectrometer used was an ultra-high throughput Witec UHTS 300 with up to 70\% throughput, designed specifically for Raman microscopy. The integration time and number of accumulations were in average 60 s and 3, respectively.

\section{Crystalline Structure and Group Theory}

Both CoFe${}_{2}$O${}_{4}$  and CuFe${}_{2}$O${}_{4}$ crystallizes according to inverse spinel-based structure. Particularly, CuFe${}_{2}$O${}_{4}$ assumes a body centered tetragonal distorted inverse-spinel structure belonging to the space group $I4_{1}/amd (D_{4h}^{19})$, in which there are four molecules per unit cell $\left( Z = 4 \right)$. In this structure Fe/Cu, Fe and O atoms occupy 8d (C${}_{2h}$), 4a (D${}_{2}$${}_{d}$) and 16h (C${}_{s}$) Wyckoff sites, respectively. On the other hand, CoFe${}_{2}$O${}_{4}$${}_{ }$ shows a face centered cubic inverse-spinel structure, belonging to the space group $Fd\overline{3}m (O_h^7)$, with eight molecules per unit cell $\left( Z = 8 \right)$, in which Fe/Co, Fe and O atoms occupy 16d (D${}_{3d}$), 8a (T${}_{d}$) and 32e (C${}_{3v}$) Wyckoff sites, respectively.

Since each structure contains 14 atoms in the primitive unit cell, there are 42 degrees of freedom and, consequently, 42 phonons are permitted for both structures. Using the Factor Group Analysis \cite{PORTO1981},  the zone-center vibrational modes distribution was decomposed in terms of the irreducible representations for both O\textit{${}_{h}$ }and D${}_{4}$\textit{${}_{h}$ } factor groups (Table \ref{table1}). Due to these site occupations, for both structures, Fe$^B$/M (M = Co or Cu) atoms do not contribute to the Raman-active phonon spectra. Thus, any Raman assignments can be attributed to (Fe)$^A$ and O ions, while the infrared spectra may be influenced by all constituent ions.

\begin{table}
  \centering
  \begin{tabular}{ccccl} \hline
{\bf Structure} & {\bf Atoms} & {\bf Sites} & {\bf Site symmetry} & {\bf Irreducible representations} \\ \hline
\multirow{8}{*}{{\bf Cubic} $\left( O_{h}^{7} \right)$} & Fe/Co & $16d$ & $D_{3d}$ & $A_{2u} \oplus  E_{u} \oplus 2F_{1u} \oplus F_{2u}$ \\
& Fe & $8a$ & $T_{d}$ & $F_{1u} \oplus F_{2g}$ \\
& O & $32e$ & $C_{3v}$ & $A_{1g} \oplus A_{2u} \oplus E_{g} \oplus E_{u} \oplus F_{1g} \oplus 2F_{1u} \oplus 2F_{2g} \oplus F_{2u}$ \\ \cline{2-5}
& & & Total & $A_{1g} \oplus 2A_{2u} \oplus E_{g} \oplus 2E_{u} \oplus F_{1g} \oplus 5F_{1u} \oplus 3F_{2g} \oplus 2F_{2u}$ \\
& & & Acoustic modes & $F_{1u}$\\
& & & IR modes & $4F_{1u}$ \\
& & & Raman modes & $A_{1g} \oplus E_{g} \oplus 3F_{2g}$ \\
& & & Silent modes & $2A_{2u} \oplus 2E_{u} \oplus F_{1g} \oplus 2F_{2u}$ \\ \hline
\multirow{10}{*}{{\bf Tetragonal} $ \left( D_{4h}^{19} \right)$ } & Fe/Cu & $8d$ & $C_{2h}$ & $A_{1u} \oplus 2A_{2u} \oplus B_{1u} \oplus 2B_{2u} \oplus 3E_{u}$ \\
& Fe & $4a$ & $D_{2d}$ & $A_{2u} \oplus B_{1g} \oplus E_{g} \oplus E_{u}$ \\
& O & $16$h & $C_{s}^{v}$ & $2A_{1g} \oplus A_{1u }\oplus A_{2g} \oplus 2A_{2u} \oplus 2B_{1g } \oplus B_{1u } \oplus B_{2g } $\\
            & & & & $\oplus 2B_{2u} \oplus3E_{g} \oplus 3E_{u}$ \\ \cline{2-5}
& & & Total & $2A_{1g} \oplus 2A_{1u }\oplus A_{2g} \oplus 5A_{2u} \oplus 3B_{1g } \oplus 2B_{1u } \oplus B_{2g } $\\
            & & & & $\oplus 4B_{2u} \oplus 4E_{g} \oplus 7E_{u}$ \\
& & & Acoustic modes & $A_{2u} \oplus E_{u}$ \\
& & & IR modes & $4A_{2u} \oplus 6E_{u}$ \\
& & & Raman modes & $2A_{1g} \oplus 3B_{1g } \oplus B_{2g }\oplus 4E_{g}$ \\
& & & Silent modes & $2A_{1u }\oplus A_{2g} \oplus 2B_{1u } \oplus 4B_{2u}$ \\ \hline
\end{tabular}
  \caption{ Factor group analysis from CoFe${}_{2}$O${}_{4}$ (cubic) and CuFe${}_{2}$O${}_{4 }$(tetragonal)}\label{table1}
\end{table}

CuFe$_2$O$_4$ and CoFe$_2$O$_4$ spinel Raman and infrared spectra patterns were also analyzed according to the quasi-molecular description, using the internal modes for the FeO$_4$ tetrahedron \cite{Verble1974}. So, in this case the structures was described as formed by two different sub-lattice groups:  Fe$^B$/M atoms and (FeO$_{4}$)$^{5-}$ tetrahedron. Tables \ref{table2} and \ref{table3} summarize the FeO${}_{4}$ group vibrations obtained by the correlation diagrams for cubic and tetragonal structures, respectively.

\begin{table}
  \centering
  \includegraphics[width=10cm,angle=0]{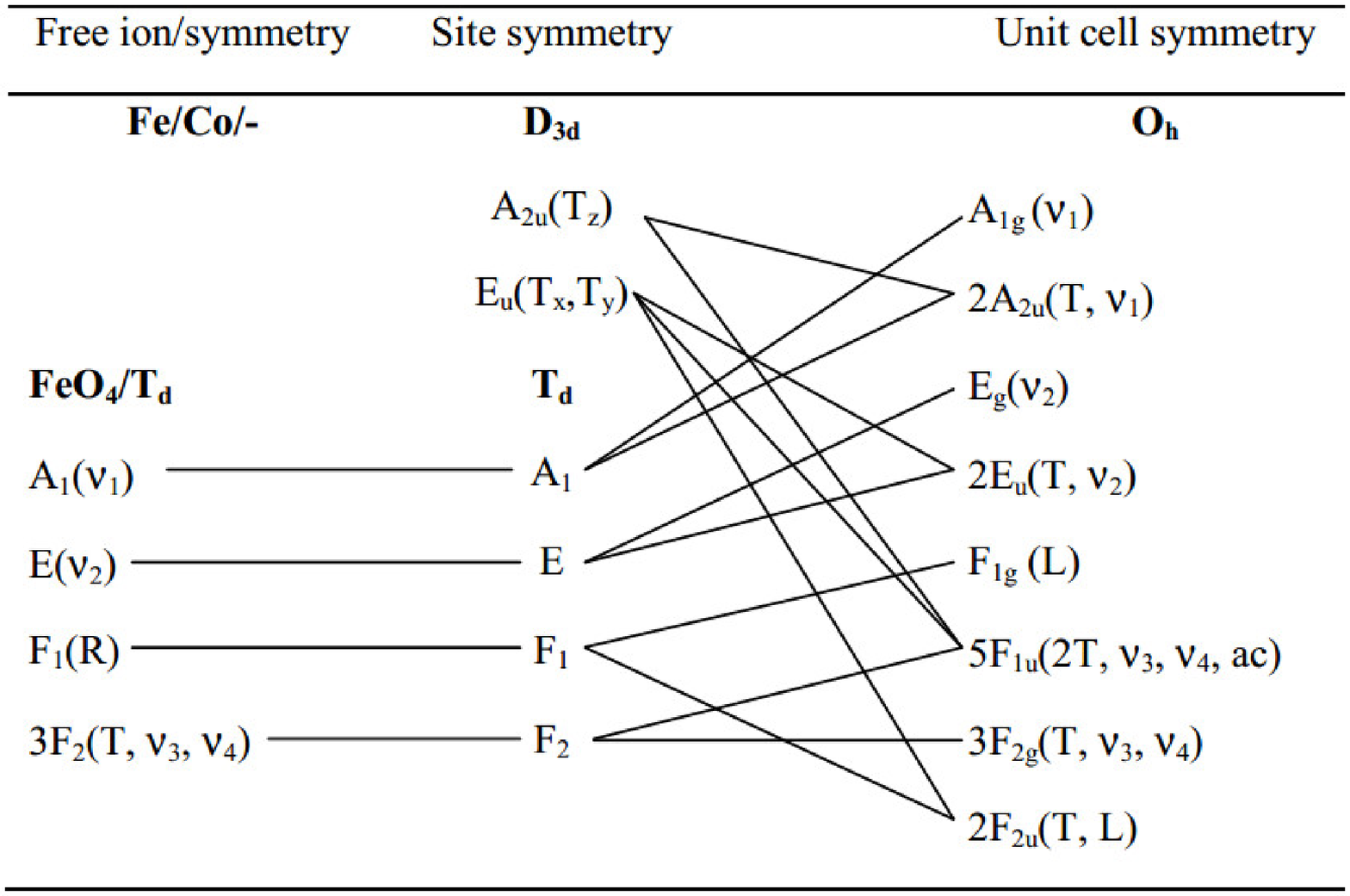}
  \caption{Correlation charts of the phonon symmetry for AFe$_{2}$O$_{4}$ in the $ \left( O_{h}^{7} \right) $ cubic structure.}\label{table2}
\end{table}

\begin{table}
  \centering
  \includegraphics[width=10cm, angle=0]{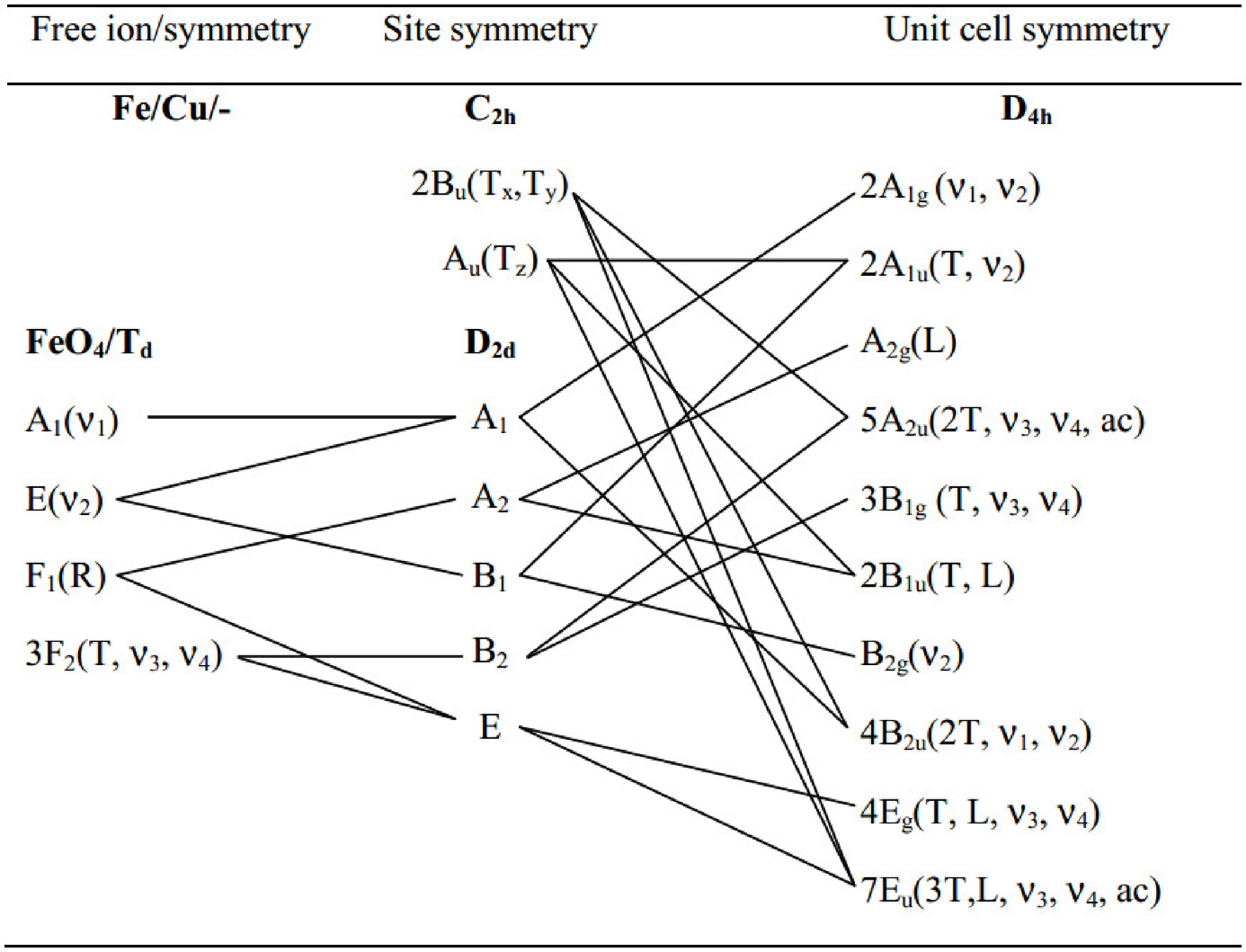}
  \caption{Correlation charts of the phonon symmetry for AFe$_{2}$O$_{4}$ in the $ \left( D_{4h}^{19} \right)$ tetragonal phase.}\label{table3}
\end{table}

According to the FeO$_4$ symmetry, $\nu_1 \left( A_1 \right) $ and $\nu_3 \left( F_2 \right) $ are assigned to the Fe--O symmetric and asymmetric stretchings, respectively; while $\nu_2 \left( E \right) $ and $\nu_1 \left( F_2 \right) $ vibrational modes are attributed to the symmetric and asymmetric Fe-O bendings. The $T \left( F_2 \right) $ vibrational mode is assigned to the FeO$_4$ tetrahedron translation motion. Finally, the Raman and infrared modes for both cubic and tetragonal inverse spinels, disregarding the acoustic and silent modes, are shown  in Table \ref{table4}.
\begin{table}
  \centering
  \begin{tabular}{ccl}
    \hline
    {\bf Spinel structure} & {\bf Activity }& {\bf Assignments }\\ \hline
    \multirow{3}{*}{$D_{4h}^{19}$(CuFe$_2$O$_4$)} & Raman & $2T ( B_{1g}\oplus E_g) +L(E_g)+\nu_1(A_{1g})+ 2\nu_2(A_{1g}\oplus B_{2g})+2\nu_3(B_{1g}\oplus E_g)$+\\
    & & $2\nu_4(B_{1g}\oplus E_g)$ \\
    & IR & $5T (2A_{2u}\oplus 3E_u) +L(E_u)+2\nu_3(A_{2u}\oplus E_u)+2\nu_4(A_{2u}\oplus E_u) $ \\ \hline
   \multirow{2}{*}{$O_{h}^{7}$ (CoFe$_2$O$_4$) } & Raman & $T(F_{2g}) \oplus \nu_1 (A_{1g}) \oplus \nu_2 (E_g) \oplus \nu_3 (F_{2g}) \oplus \nu_4 (F_{2g})$ \\
     & IR & $2T (2F_{1u}) \oplus \nu_3 (F_{1u}) \oplus \nu_4 (F_{1u})$ \\
   \hline
  \end{tabular}
  \caption{Raman and Infrared modes corresponding to the cubic and tetragonal inverse spinels}\label{table4}
\end{table}

\section{Results and Discussions}

\subsection{Structural properties}

\begin{figure}
  \centering
  \includegraphics[width=15cm]{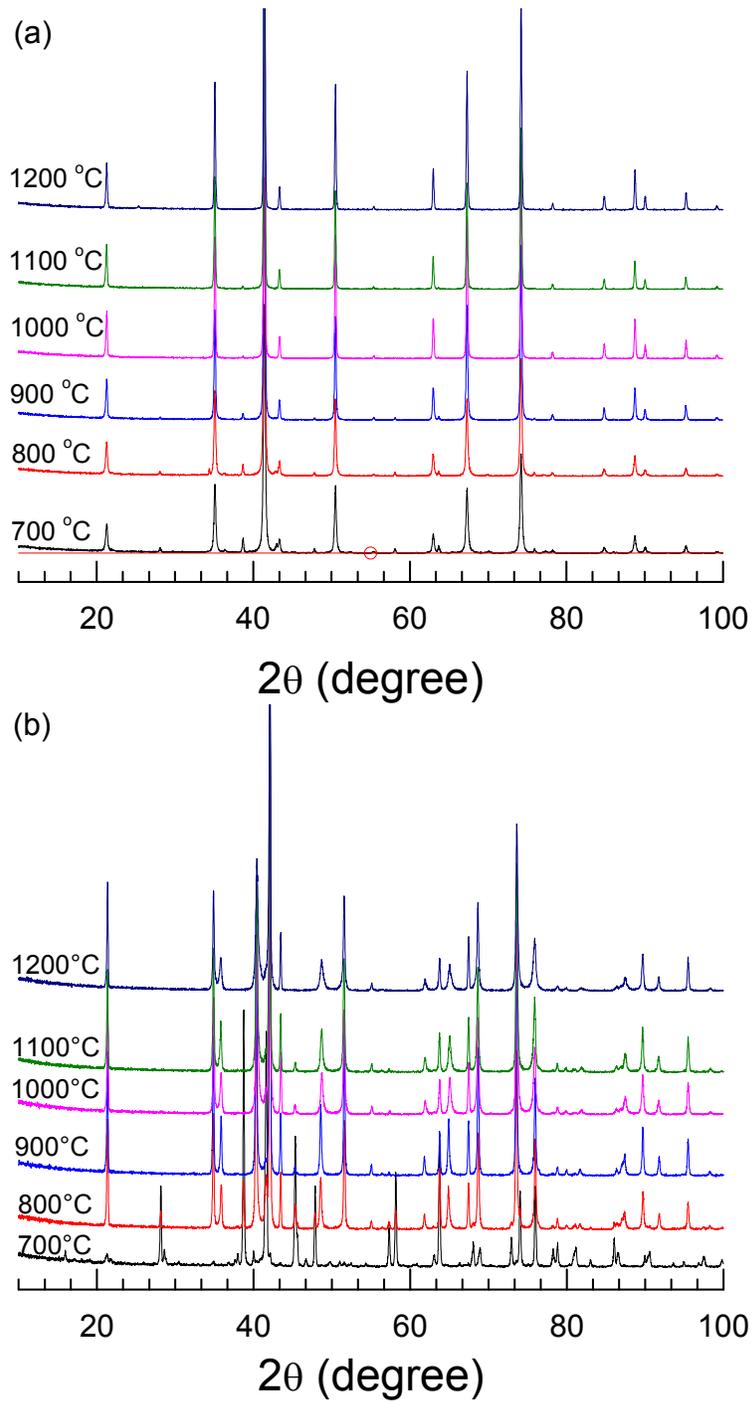}\\
  \caption{XRD patterns as function of the annealing temperature for (a) CoFe$_{2}$O$_{4}$ and (b) CuFe${}_{2}$O${}_{4}$ spinels.}
  \label{xrd_pure}
\end{figure}

\begin{figure}
  \centering
  \includegraphics[width=15cm]{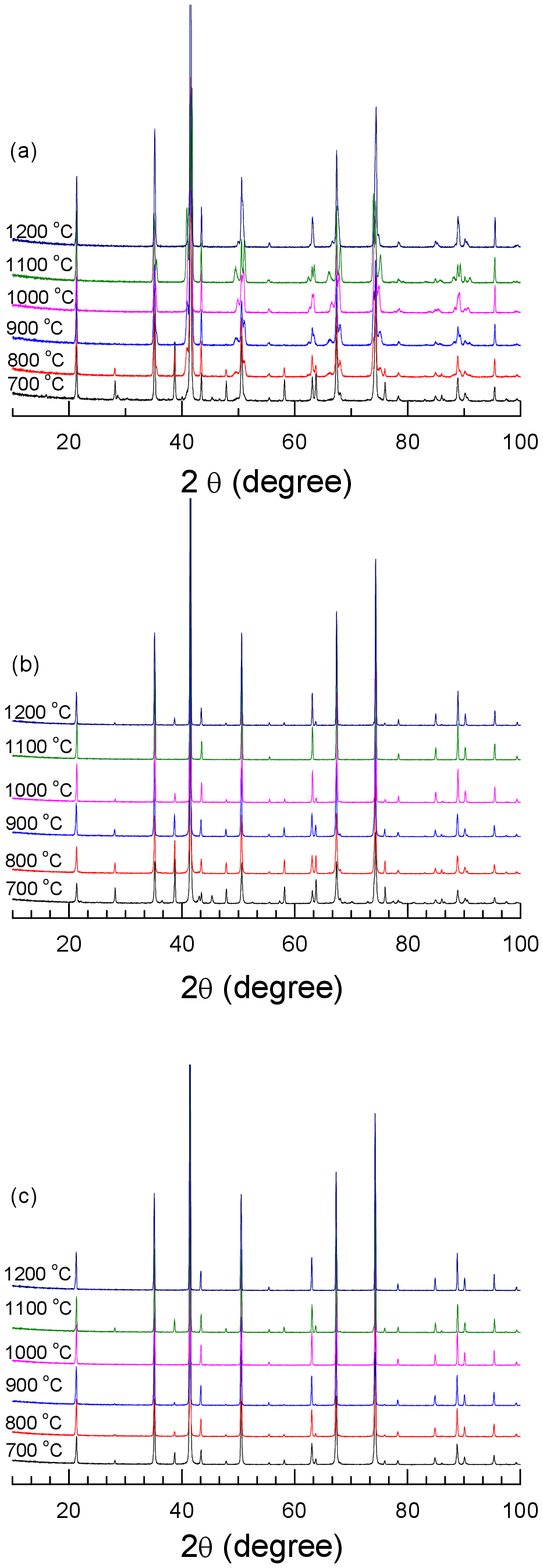}\\
  \caption{XRD patterns as function of the annealing temperature for (a) Co${}_{0.25}$Cu${}_{0.75}$Fe${}_{2}$O${}_{4}$. (b) Co${}_{0.50}$Cu${}_{0.50}$Fe${}_{2}$O${}_{4}$ and (c) Co${}_{0.75}$Cu${}_{0.25}$Fe${}_{2}$O${}_{4}$ spinels.}
  \label{xrd_mixed}
\end{figure}

\begin{figure}
  \centering
  \includegraphics[width=10cm, angle=0]{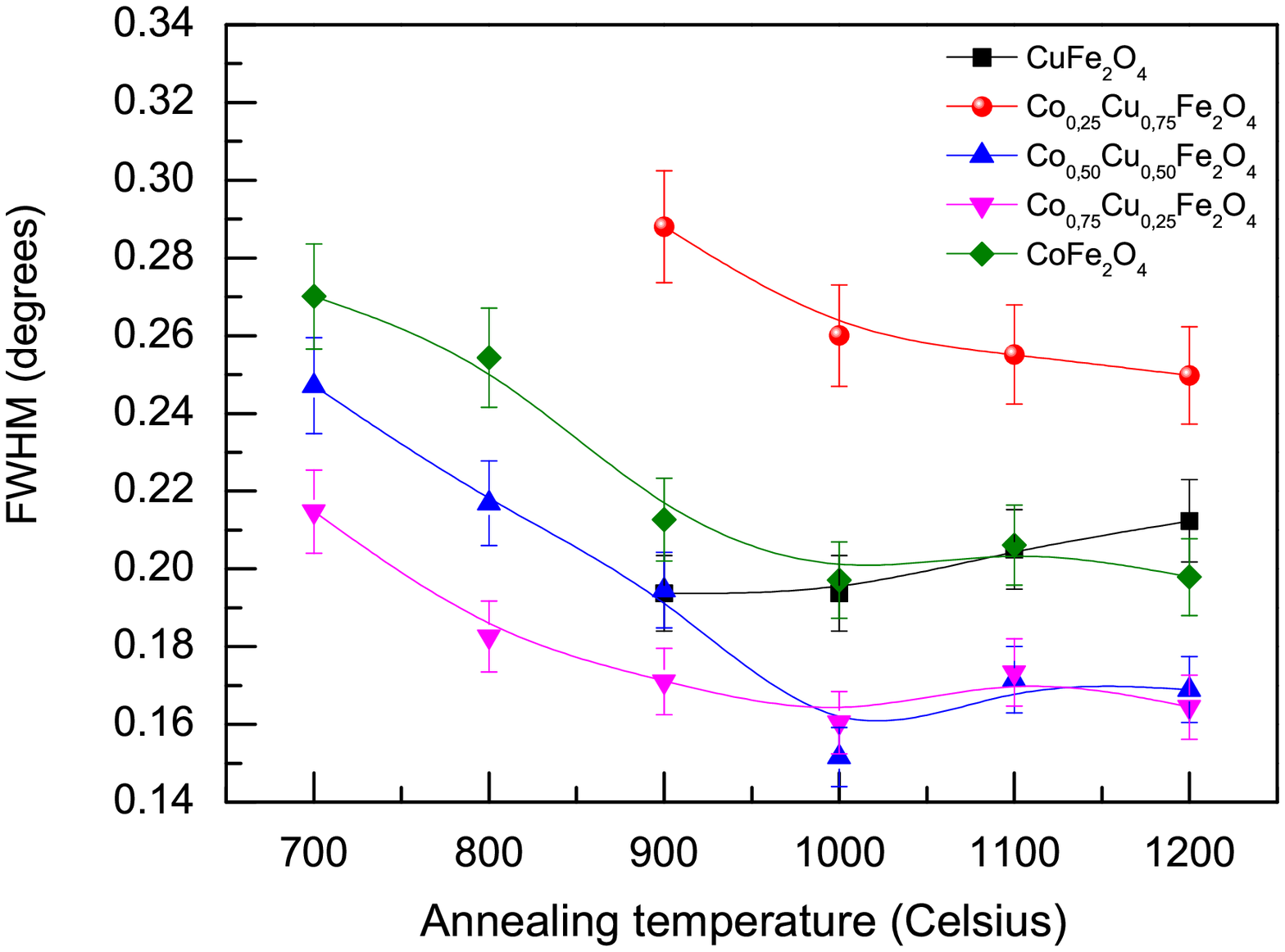}\\
  \caption{FWHM as function of the annealing temperature for the Co$_{1-x}$Cu$_{x}$Fe$_{2}$O$_{4}$ ($x = 1.00, 0.75, 0.50, 0.25, 0.00$) spinels. The lines are guides for the eyes.}\label{FWHM}
\end{figure}

\begin{figure}
  \centering
  \includegraphics[width=10cm, angle=0]{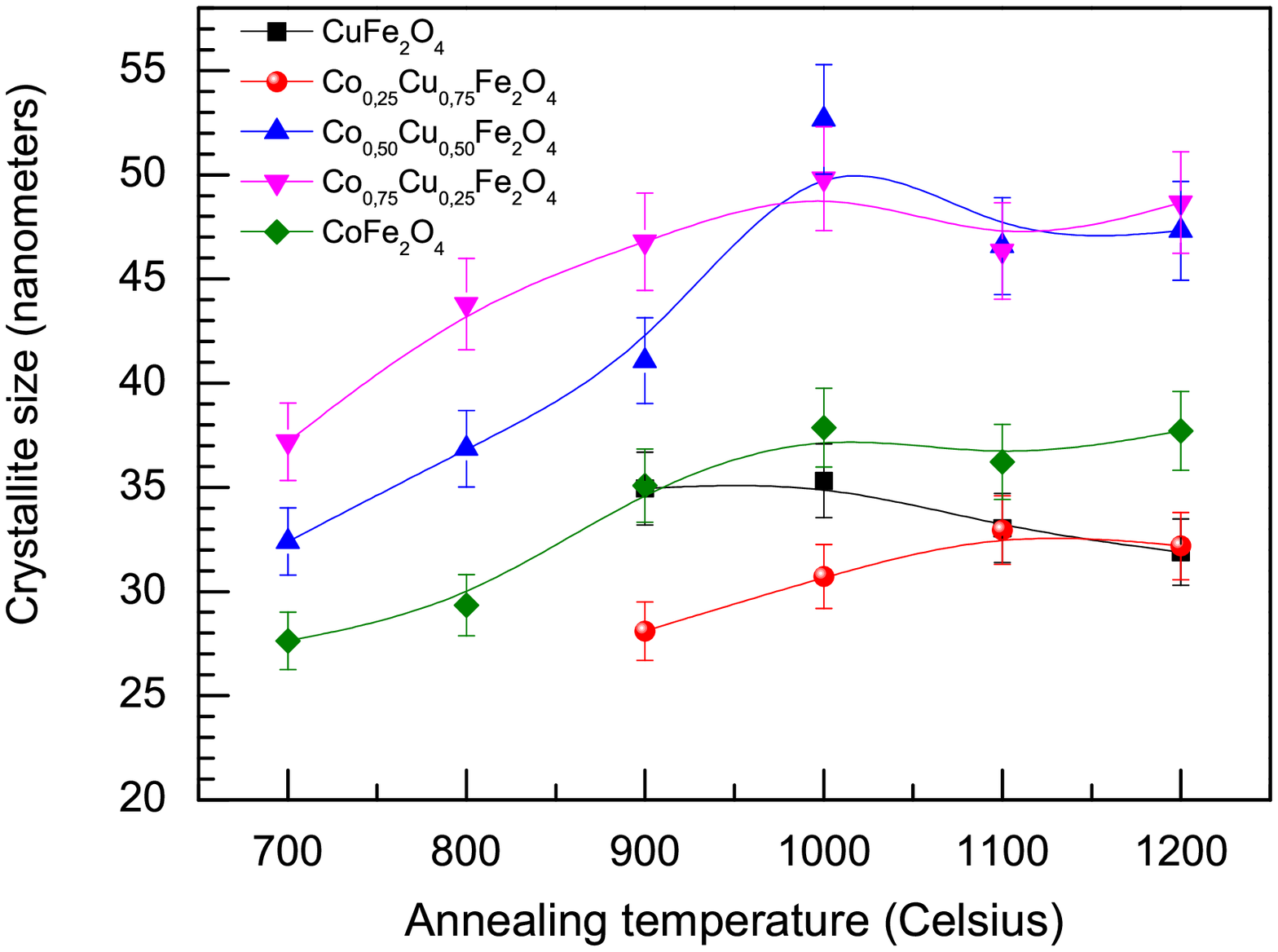}\\
  \caption{Crystallite size as function of the ternary thermal treatment for the Co$_{1-x}$Cu$_{x}$Fe$_{2}$O$_{4}$ ($x = 1.00, 0.75, 0.50, 0.25, 0.00$) spinels. The lines are guides for the eyes.}\label{size}
\end{figure}

Figures \ref{xrd_pure} and \ref{xrd_mixed} present the XRD patterns for the several Co$_{1-x}$Cu$_{x}$Fe$_{2}$O$_{4}$ spinel powders as a function of the annealing temperature. The results confirm the formation of the single phase spinel, as indicated by the JCPDS 00-034-0425 (tetragonal CuFe$_2$O$_4$) and JCPDS 00-022-1086 (cubic CoFe$_2$O$_4$) patterns, excepted for those annealed at 700 and 800 $^\circ$C Cu-based pure spinels that contain $\alpha$-Fe$_2$O$_4$ rhombohedral (JCPDS 00-013-0534). It is important to point out that $\alpha$-Fe$_2$O$_4$ is often found as a secondary phase in spinel synthesis, as previously reported by Sun et al. \cite{SUN2007} and Mathew et al. \cite{MATHEW2004}. In addition, for the cobalt substituted ferrites, it can be noticed than even minor Co$^{2+}$ inclusions induce a cubic lattice.
The full-width at half maximum (FWHM) (monitoring the peak (400))  in CoFe$_2$O$_4$ decreases with the increase of the annealing temperature, as shown in Figure \ref{FWHM}, showing that the average crystallite size is becoming larger, while an inverse behavior is observed in CuFe$_2$O$_4$ (monitoring the peak (211)).
Besides, for Co$_{0.25}$Cu$_{0.75}$Fe$_{2}$O$_{4}$ samples, the FWHM hardly changes, except when synthesized at 900 $^\circ$C and above. This behavior could be associated to minor lattice distortion imposed by the spatial competition between octahedral and tetrahedral sites. Such fact is confirmed by Figure \ref{size}, since Co-substituted ferrites present bigger average crystallite sizes, as compared to the CuFe$_2$O$_4$ structures. Furthermore, the average crystallite size tends to enlarge with the annealing (26 up to 54 nm). Specifically, CuFe$_2$O$_4$ crystallites show mean size between 32 and 35 nm, much lower than the value proposed by Sun {\i et al.} \cite{SUN2007}, that estimated the range in between 75 - 110 nm for the ferrites. In opposition, CoFe$_2$O$_4$ crystallite sizes were in the range 28 - 37 nm, being also smaller than those reported by Gaikwad {\it et al.} \cite{GAIKWAD2011}, i.e., 60 - 80 nm, but bigger than the one obtained by Valdés-Solís {\it et al.} \cite{VALDES2007}, that was 20 nm.

\subsection{Vibrational properties}
\begin{figure}
  \centering
  \includegraphics[width=10cm, angle=270]{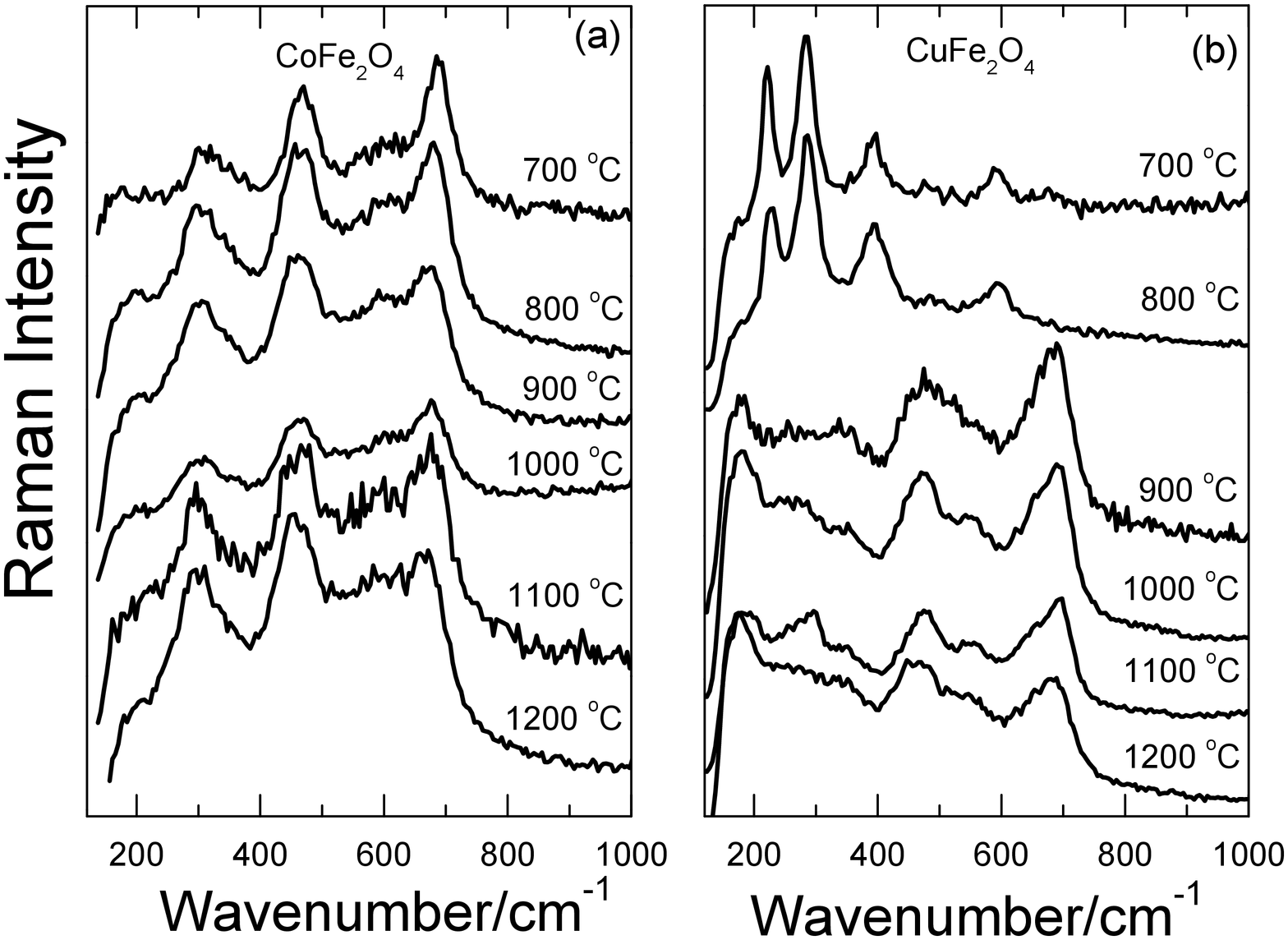}\\
  \caption{Room temperature Raman spectra for (a) CoFe$_2$O$_4$ and (b) CuFe$_2$O$_4$ investigated spinels.}\label{raman_pure}
\end{figure}

Figure \ref{raman_pure} shows the Raman spectra obtained for CoFe$_2$O$_4$ and CuFe$_2$O$_4$ pure annealed spinels. CuFe$_2$O$_4$ spinel, synthesized at 700 and 800 $^\circ$C (Figure \ref{raman_pure}b), in fact corresponds to the $\alpha$-Fe$_2$O$_4$ as confirmed by their Raman spectra and supported by the XRD results \cite{SHEBANOVA2003}. Nonetheless, for higher annealing temperatures (900 - 1200$^\circ$C),  eight characteristical vibrational modes (Table \ref{table5}) attributed to the monophasic CuFe$_2$O$_4$ structure are depicted from the spectra and such result agrees with group theory prediction. CoFe$_2$O$_4$ sample spectra (Figure \ref{raman_pure}a) exhibit only five broad bands  around 180, 315, 480, 640 and 680 cm$^{–1}$, all of them assigned to the cubic inverse-spinel ferrites \cite{Verble1974,SHEBANOVA2003,Gupta2002,Gasparov2000,Li2000,Bersani1999,deFaria1997,THIERRY1991,BOUCHERIT1991,DUNNWALD1989,GRAVES1988,DEGIORGI1987,OHTSUKA1986}
. In this structure the modes around 180 and 300 cm$^{-1}$ are assigned to the $T(F_{2g})$ and $\nu_2(E_g)$ modes, respectively. Also, the modes $\nu_4(F_{2g})$, $\nu_3(F_{2g})$ and $\nu_1(A_{1g})$ are observed at 460, 590 and 680 cm$^{-1}$, respectively.

\begin{figure}
  \centering
  \includegraphics[width=10cm, angle=270]{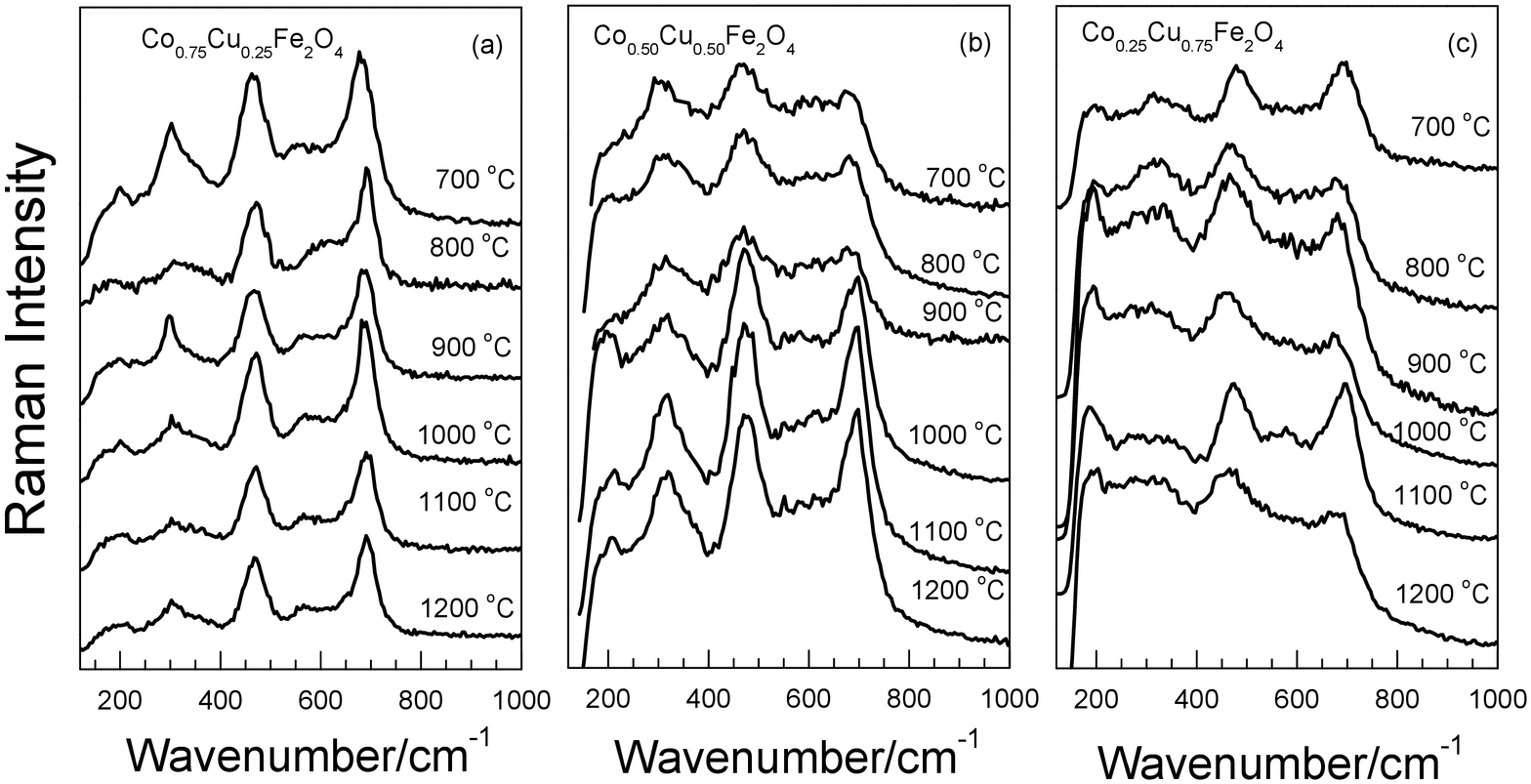}\\
  \caption{Room temperature Raman spectra for (a) Co$_{0.75}$Cu$_{0.25}$Fe$_2$O$_4$, (b) Co$_{0.50}$Cu$_{0.50}$Fe$_2$O$_4$ and (c) Co$_{0.25}$Cu$_{0.75}$Fe$_2$O$_4$ spinels.}\label{raman_mixed}
\end{figure}

Table \ref{table5} summarizes the symmetry assignments to the observed Raman-active modes of the several spinel samples investigated in this work. The wavenumber of the peaks indicated were found by fitting Raman spectra with Lorentzian functions. In general, the phonon spectral positions are essentially independent of the annealing temperature for all compounds.
\begin{table}
  \centering
  \begin{tabular}{cccccccccc} \hline
\multirow{2}{*}{{\bf Compound}} & \multirow{2}{*}{{\bf modes}} & \multicolumn{6}{c}{{\bf Annealing temperature} ($^\circ$C) }& \multirow{2}{*}{{\bf Assignment }}\\ \cline{3-8}
& & 700 & 800 & 900 & 1000 & 110 & 1200 &  \\ \hline
\multirow{5}{*}{{\bf CoFe$_2$O$_4$}} & 1 & 171 & 170 & 169 & 169 & 179 & 188 & T(F${}_{2g}$) \\
& 2 & 315 & 300 & 295 & 294 & 296 & 297 & $\nu$${}_{2}$(E${}_{g}$) \\
& 3 & 468 & 462 & 459 & 460 & 458 & 456 & $\nu$${}_{4}$(F${}_{2}$${}_{g}$) \\
& 4 & 599 & 595 & 592 & 593 & 596 & 588 & $\nu$${}_{3}$(F${}_{2}$${}_{g}$) \\
& 5 & 687 & 682 & 678 & 679 & 680 & 672 & $\nu$${}_{1}$(A${}_{1}$${}_{g}$) \\ \hline
\multirow{8}{*}{{\bf Co$_{0.75}$Cu$_{0.25}$Fe$_{2}$O$_{4}$}} & 1 & 192 & 183 & 193 & 189 & 188 & 192 & T \\
& 2 & 300 & 308 & 298 & 303 & 309 & 303 & $\nu$${}_{2}$ \\
& 3 & 355 & 355 & $-$ & 359 & $-$ & 363 & $\nu$${}_{4}$ \\
& 4 & 464 & 469 & 466 & 468 & 468 & 468 & $\nu$${}_{4}$ \\
& 5 & 552 & $-$ & 563 & 569 & 575 & 570 & $\nu$${}_{3}$ \\
& 6 & 608 & 602 & 619 & 626 & $-$ & $-$ & $\nu$${}_{3}$ \\
& 7 & $-$ & 659 & $-$ & $-$ & 657 & 641 & $\nu$${}_{3}$ \\
& 8 & 681 & 693 & 688 & 690 & 692 & 692 & $\nu$${}_{1}$ \\ \hline
\multirow{8}{*}{{\bf Co${}_{0.50}$Cu${}_{0.50}$Fe${}_{2}$O${}_{4}$}} & 1 & 196 & 189 & 203 & 180 & 188 & 187 & T \\
& 2 & $-$ & $-$ & $-$ & 239 & 239 & 245 & T/L \\
& 3 & 303 & 300 & 312 & 309 & 311 & 313 & $\nu$${}_{2}$ \\
& 4 & $-$ & $-$ & $-$ & 379 & 374 & 368 & $\nu$${}_{4}$ \\
& 5 & 467 & 471 & 469 & 472 & 470 & 473 & $\nu$${}_{4}$ \\
& 6 & 610 & 604 & 605 & 576 & 573 & 592 & $\nu$${}_{3}$ \\
& 7 & $-$ & $-$ & $-$ & 678 & 662 & 672 & $\nu$${}_{3}$ \\
& 8 & 688 & 689 & 687 & 702 & 696 & 698 & $\nu$${}_{1}$ \\ \hline
\multirow{8}{*}{{\bf Co${}_{0.25}$Cu${}_{0.75}$Fe${}_{2}$O${}_{4}$}} & 1 & 171 & 170 & 170 & 172 & 174 & 177 & T \\
& 2 & 197 & 194 & 194 & 198 & 200 & 214 & T/L \\
& 3 & 238 & 230 & 239 & 245 & 268 & 266 & $\nu$${}_{2}$ \\
& 4 & 320 & 313 & 322 & 321 & 346 & 331 & $\nu$${}_{4}$ \\
& 5 & 482 & 468 & 466 & 460 & 475 & 463 & $\nu$${}_{4}$ \\
& 6 & 586 & 582 & 597 & 588 & 577 & 587 & $\nu$${}_{3}$ \\
& 7 & 685 & 651 & 682 & 673 & 667 & 659 & $\nu$${}_{3}$ \\
& 8 & 712 & 693 & 711 & 702 & 701 & 694 & $\nu$${}_{1}$ \\ \hline
\multirow{8}{*}{{\bf CuFe${}_{2}$O${}_{4}$}} & 1 & $-$ & $-$ & 177 & 171 & 177 & 168 & T \\
& 2 & $-$  & $-$  & 250 & 244 & 245 & 211 & T/L \\
& 3 & $-$  & $-$  & $-$ & $-$ & 294 & 271 & $\nu$${}_{2}$ \\
& 4 & $-$  & $-$  & 343 & 348 & 348 & 346 & $\nu$${}_{4}$ \\
& 5 & $-$  & $-$  & 467 & 469 & 474 & 462 & $\nu$${}_{4}$ \\
& 6 & $-$  & $-$  & 532 & 554 & 554 & 549 & $\nu$${}_{3}$ \\
& 7 & $-$  & $-$  & 663 & 658 & 655 & 656 & $\nu$${}_{3}$ \\
& 8 & $-$  & $-$  & 694 & 696 & 696 & 692 & $\nu$${}_{1}$ \\ \hline
\end{tabular}
  \caption{Wavenumbers (in cm$^{-1}$) and Raman modes assignments for Co$_{1-x}$Cu$_x$Fe$_2$O$_4$ $(x = 1.00, 0.75, 0.50, 0.25, 0.00)$ spinels.}\label{table5}
\end{table}

Figure \ref{raman_mixed} shows the Raman spectra obtained from all mixed spinels ($x=0.75\text{, } 0.5\text{ and }0.25$). Although the XRD patterns suggest that all Cobalt-doped samples' structures are cubic, the Raman spectra reveals that Co$_{0.25}$Cu$_{0.25}$Fe$_2$O$_4$, Co$_{0.50}$Cu$_{0.50}$Fe$_2$O$_4$ and Co$_{0.75}$Cu$_{0.25}$Fe$_2$O$_4$ locally assume tetragonal structures. Such discrepancy is explained  localized nature of Raman spectroscopy in contrast to the averaged one of XRD. Besides, no significative wavenumber mode dependence was observed with Cobalt substitution.

\begin{figure}
  \centering
  \includegraphics[width=10cm, angle=270]{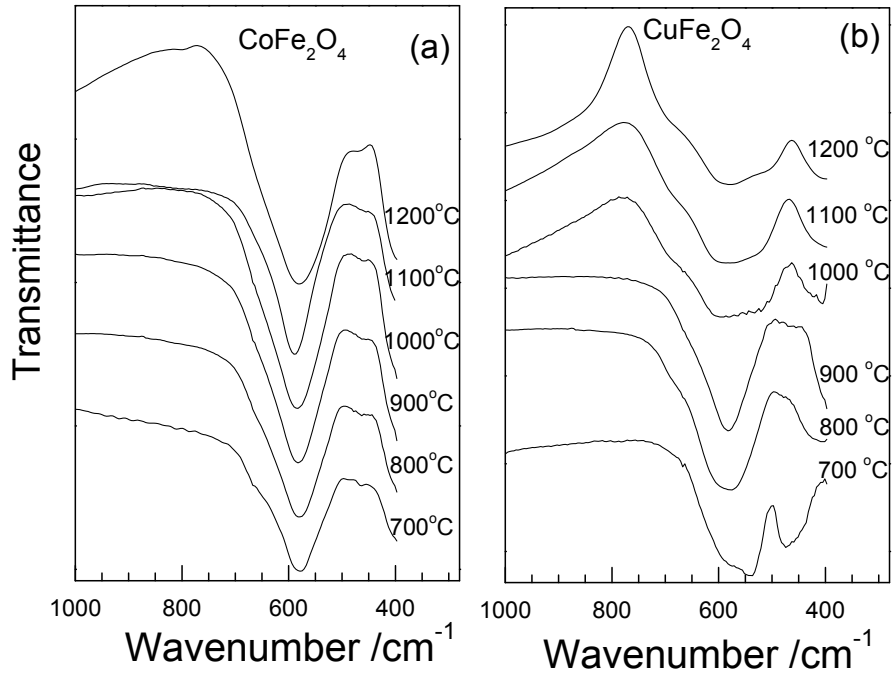}\\
  \caption{Room temperature IR spectra of (a) CoFe$_2$O$_4$ and (b) CuFe$_2$O$_4$ investigated spinels.}\label{ir_pure}
\end{figure}

Figure \ref{ir_pure} shows that the room temperature mid IR spectra observed for CuFe$_2$O$_4$ and CoFe$_2$O$_4$ spinels are in good agreement with those reported by previous works. \cite{Hashim2012,Eshraghi2011,Cedeno-Mattei2009,Zhao2008,SELVAN2003,Gillot1997,Gillot1997b,Gillot1997c,WALDRON1955}. According to Selvan et al. \cite{SELVAN2003} and Dey et al. \cite{DEY2003} CuFe$_2$O$_4$ powders have two characteristical vibrational modes about 600 and 400 cm$^{-1}$, which are attributed to the tetrahedral and octahedral M$^{2+}$ cations, respectively. In this work, such stretching modes are detected around 474 - 397 cm$^{-1}$ [$\nu$(Fe-O)$_B$], 531- 559 cm$^{-1}$ [$\nu$(Fe-O)$_A$] and 605 - 672 cm$^{-1}$ [$\nu$(Cu-O)$_A$], in which both Fe-O modes are assigned to the $F_{1u}$ symmetry (Figure \ref{ir_pure}). The higher wavenumber values for the A-site cations results from the shortening M$^{2+}$-O$^{2-}$ bond length at such symmetries \cite{COSTA2011}.

Once again, the $\alpha$-Fe$_2$O$_4$ phase is depicted from the spectra of CuFe$_2$O$_4$ samples annealed at 700 $^\circ$C and 900 $^\circ$C, as indicated by the strong absorption band at 475 cm$^{-1}$ [$\nu$(Fe-O)] \cite{GOTIC2007}. Finally, the CuFe$_2$O$_4$ samples obtained at 1000, 1100 and 1200 $^\circ$C present a significant tetragonal distortion (Jahn-Teller distortion), corroborating to the Raman and XRD data.

Considering the CoFe$_2$O$_4$ sample, similar vibrational modes are observed around 399 - 400 [$\nu$(Fe-O)$_B$], 464 - 541 [$\nu$(Fe-O)$_A$] and 576 - 583 [$\nu$(Co-O)$_A$] (Figure \ref{ir_pure}a). The displacement of the (Co-O)$_A$ modes to lower wavenumbers, as compared to the (Cu-O)$_A$ one, results from the smaller cobalt ionic radius, generating a larger vibrational energy \cite{POPMINTCHEV2009,SINGHAL2011,ZHANG2012}.

\begin{figure}
  \centering
  \includegraphics[width=10cm, angle=270]{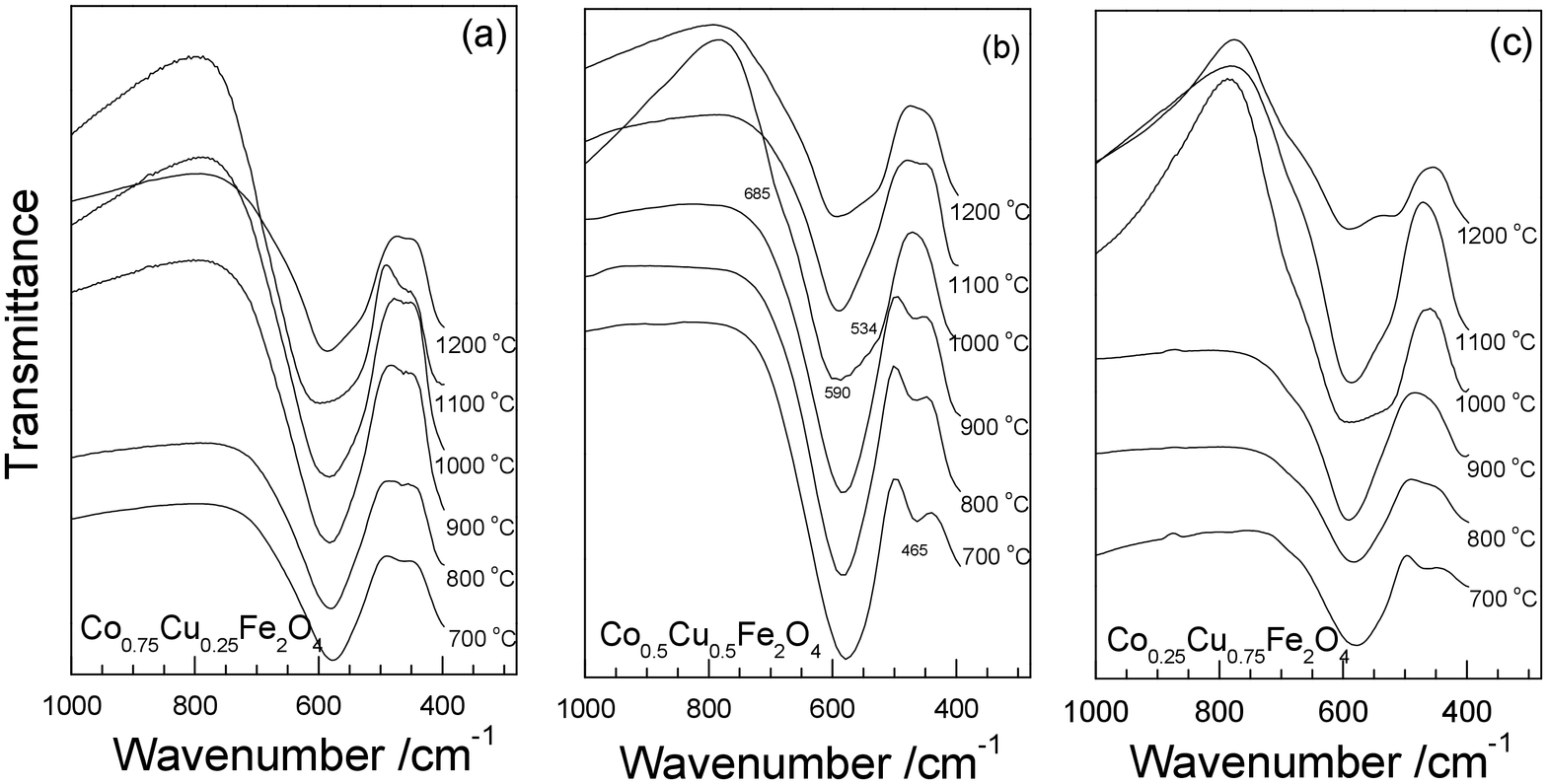}\\
  \caption{Room temperature IR spectra of (a) Co$_{0.75}$Cu$_{0.25}$Fe$_2$O$_4$, (b) Co$_{0.50}$Cu$_{0.50}$Fe$_2$O$_4$ and (c) Co$_{0.25}$Cu$_{0.75}$Fe$_2$O$_4$ spinels.}\label{ir_mixed}
\end{figure}

The Co-substituted CuFe$_2$O$_4$ IR spectra (Figure \ref{ir_mixed}) show a continuous band narrowing as function of cobalt contents into the lattice. Nevertheless, the same tetrahedral and octahedral vibrational modes are assigned.


\subsection{Conclusions}

In this work we have synthesized nanometric Co${}_{1-x}$Cu${}_{x}$Fe${}_{2}$O${}_{4}$ spinels using a polymeric  precursors method precursors. The crystalline structures were investigated by X-ray powder diffraction, which confirms that the pure CoFe${}_{2}$O${}_{4}$ and CuFe${}_{2}$O${}_{4}$ samples have cubic and tetragonal inverted spinel structures, respectively. The crystallite sizes of the mixed samples are bigger than those of the pure samples. The synthesized samples were investigated by Raman and infrared spectroscopies. This analysis showed that there is a tetragonal local distortion into the mixed sample. A complete mode assignment for all the samples was performed based on the internal vibrations of the FeO${}_{4}$ molecular group.

\acknowledgement
The authors are grateful to the Brazilian funding agencies CAPES, CNPq, and FAPEMA.

\bibliography{cof2}

\end{document}